\documentclass{article}

\usepackage{PRIMEarxiv}

\usepackage[utf8]{inputenc} 
\usepackage[T1]{fontenc}    
\usepackage{hyperref}       
\usepackage{url}            
\usepackage{booktabs}       
\usepackage{amsfonts}       
\usepackage{amsmath}        
\usepackage{nicefrac}       
\usepackage{microtype}      
\usepackage{lipsum}
\usepackage{fancyhdr}       
\usepackage{graphicx}       
\graphicspath{{media/}}     

\title{Advancing data-driven broadband seismic wavefield simulation with multi-conditional diffusion model
}

\author{
  Zhengfa Bi\\
  Lawrence Berkeley National Laboratory \\
  \texttt{zfbi@lbl.gov} \\
   \And
  Nori Nakata \\
  Lawrence Berkeley National Laboratory \\
  \texttt{nnakata@lbl.gov} \\
   \And
  Rie Nakata \\
  Lawrence Berkeley National Laboratory \\
  \texttt{rnakata@lbl.gov} \\
   \And
  Pu Ren \\
  Lawrence Berkeley National Laboratory \\
  \texttt{pren@lbl.gov} \\
  \And
  Xinming Wu \\
  University of Science and Technology of China \\
  \texttt{xinmwu@ustc.edu.cn} \\
  \And
  Michael W. Mahoney \\
  University of California, Berkeley \\
  \texttt{mmahoney@stat.berkeley.edu} \\
}

\begin{document}
\maketitle

\begin{abstract}
    Sparse distributions of seismic sensors and sources pose challenges for subsurface imaging, source characterization, and ground motion modeling. While large-N arrays have shown the potential of dense observational data, their deployment over extensive areas is constrained by economic and logistical limitations. Numerical simulations offer an alternative, but modeling realistic wavefields remains computationally expensive. To address these challenges, we develop a multi-conditional diffusion transformer for generating seismic wavefields without requiring prior geological knowledge. Our method produces high-resolution wavefields that accurately capture both amplitude and phase information across diverse source and station configurations. The model first generates amplitude spectra conditioned on input attributes and subsequently refines wavefields through iterative phase optimization. We validate our approach using data from the Geysers geothermal field, demonstrating the generation of wavefields with spatial continuity and fidelity in both spectral amplitude and phase. These synthesized wavefields hold promise for advancing structural imaging and source characterization in seismology.
\end{abstract}

\keywords{Deep learning \and Diffusion model \and Seismic simulation}

\section{Introduction}
\label{sxn:introduction}

    Seismic observations with denser source and receiver geometries have become essential for high-resolution subsurface imaging, better event and phase detectability, and more accurate source characterization and hazard assessment.
    Although the number of seismometers in the world keeps increasing (Figure II in \cite{2019booknakata}), and we have lowered the magnitude completeness, the limited density of receivers causes aliasing of wavefields and limits the resolution of subsurface imaging.
    Source detection, location, and characterization also suffers by this sparseness because of, for example, the less stacking power over receivers, the lack of enough reference events for relative locations, and the aperture of the seismic network.
    The limited number of earthquakes also increases the uncertainty of ground motion modeling.
    To overcome these challenges, here we propose a data-driven multi-conditional diffusion transformer to generate seismic wavefields at arbitrary receiver and source locations.

    Hoping to address these challenges, Machine Learning (ML) has emerged as a promising path to synthesize seismic wavefields~\cite{mousavi2020machine,mousavi2022deep}.
    The ML approach uses direct observational data, while minimizing dependence on velocity models and source parameter knowledge. 
    Due to the data-driven procedure, ML approaches potentially offer flexibilities to simulate complex, non-linear wave propagation phenomena, which might be difficult to capture with physics-based numerical methods. 
    \cite{ni2024wavefield} propose a sequence-to-sequence approach to generate ambient-noise waveforms based on DAS recordings, but their results are limited to low frequencies. 
    \cite{liu2024generative} use diffusion models to expand wavefields from sparsely sampled data within predefined interpolation grids. 
    Their approach relies on prior seismic data, and it struggles with irregularly distributed or spatially aliased datasets for high frequencies (e.g., $>1$ Hz).
    Bridging these gaps demands conditional generative models that can capture high-frequency, high-wavenumber components from such datasets without requiring prior data as input.
    

    Generative models have shown potential to generate realistic waveforms. 
    However, Generative Adversarial Networks (GANs) have limitations, e.g., due to mode collapse~\cite{wang2021seismogen,esfahani2023tfcgan,chen2024deep}, where the generated outputs lack diversity and resemble the training data excessively. 
    Similarly, conditional Variational Autoencoders (VAEs) have been successfully applied to waveform generation~\cite{li2020seismic,ren2024learning}, achieving reasonable performance in capturing overall temporal-frequency structures; but they tend to produce overly-smoothed wavefields, limiting their ability to  preserve fine-grained temporal and spectral details~\cite{gao2020zero,niu2020lstm,naiman2024iclr}.
    Therefore, in spite of the fact that \cite{ren2024learning} produce reliable spatial distribution of ground motion parameters as well as arrival times for P and S waves, their detailed wavefields (including phase) do to match very well to the earthquake records.
    Other recent advances include a neural-operator-based approach for rapid waveform modeling and inversion to improve efficiency over traditional
    seismic analysis methods~\cite{yang2023rapid}.
    While effective for approximating functional mappings, this approach requires predefined physical models or prior knowledge, which constrains their generalization to complex and unknown geological environments.
    Despite their encouraging results, these models primarily target spectra or ground motions, or require physical knowledge, and they often fail to reconstruct realistic wavefields containing both phase and amplitudes.
    
    Recently, diffusion models (DMs)~\cite{kingma2021variational,rombach2022high} have emerged as powerful generative frameworks, capable of producing high-fidelity synthetic data by progressively adding and reversing noise, resulting in outputs with both quality and diversity~\cite{ho2022cascaded}.
    DMs were originally developed for image generation and have demonstrated their ability to capture detailed spatial patterns in real-world data~\cite{croitoru2023diffusion}.
    DMs excel in capturing complex patterns and structures, making them well-suited for seismic waveform generation to simulate wavefields that closely mimic real-world observations while preserving physical characteristics and stochastic variability.

\begin{figure*}[t!]
          \centering
          \includegraphics[width=0.9\textwidth]{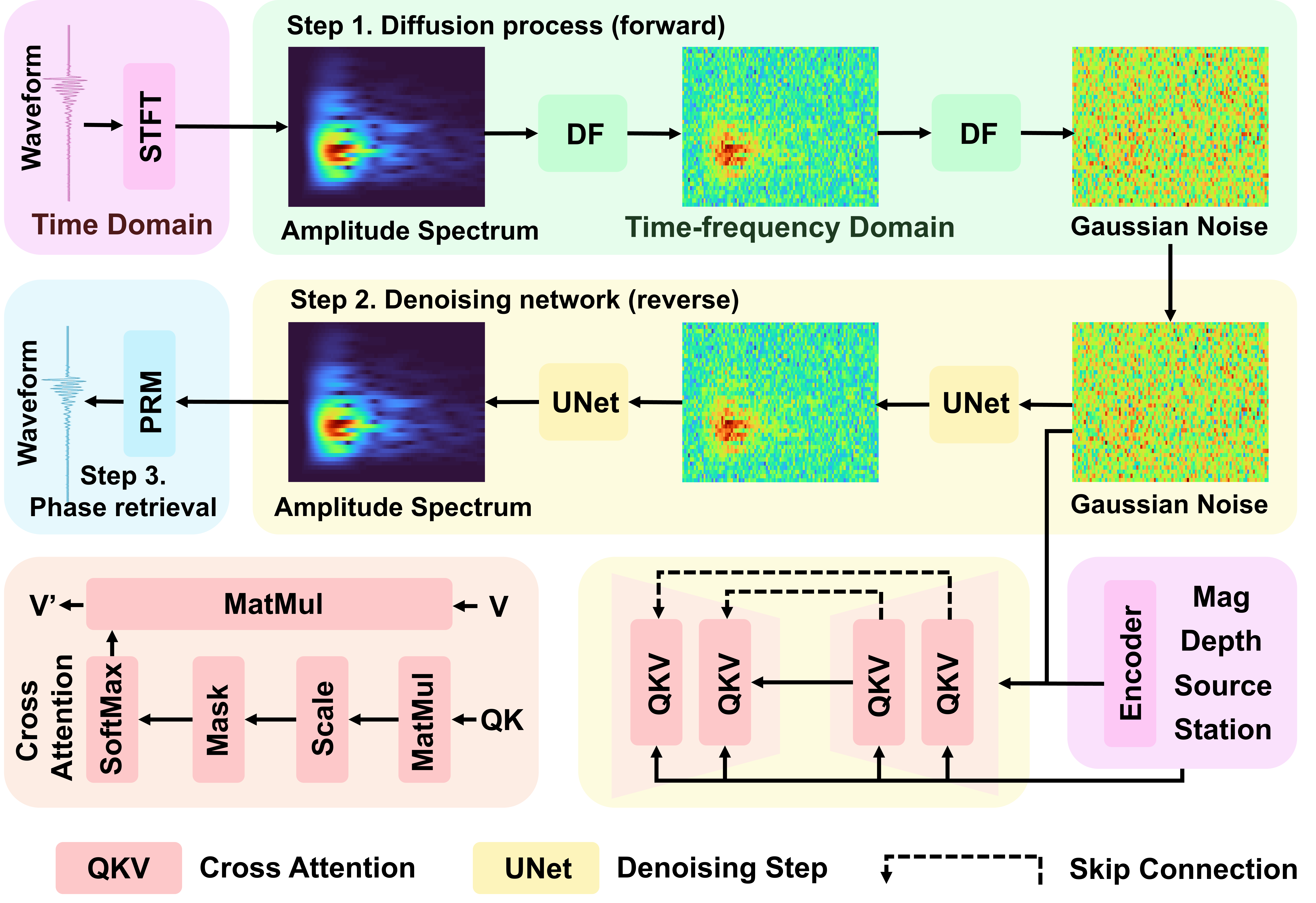}
          \caption{
          Schematic overview of CGM-Wave.
          This framework incorporates a diffusion process, noise manipulation, and a U-shaped transformer architecture, conditioned on station locations, source coordinates, and magnitude.
          The adaptive Phase Retrieval Method (PRM) is designed to reconstruct the phase spectrum and produce the waveform.
          }
          \label{fig:fwork}
    \end{figure*}

    In this paper, we present CGM-Wave, a diffusion-based conditional generative modeling approach, which contains diffusion processes and U-shaped transformers, for wavefield generation with input parameters such as source and receiver locations and source parameters (Figure~\ref{fig:fwork}). 
    By employing the Short-Time Fourier Transform (STFT) to project seismic data into the time-frequency domain, CGM-Wave enables the precise generation of spectra that faithfully represent the temporal and spectral dynamics of real-world wavefields.
    CGM-Wave incorporates a stochastic differential equation within a Markovian framework, to represent the stochastic characteristics of seismic data caused by subsurface heterogeneities and complex source mechanisms.
    A cross-attention mechanism provides fine-grained control over the generative process, allowing one to introduce domain-specific constraints and guide synthesis towards more realistic waveforms.
    An adaptive phase retrieval method (PRM) integrates seamlessly into our deep generative pipeline, promoting physically-plausible and noise-robust phase reconstructions for accurate synthesis. 
    
    Our goal is to learn representations of wave physics and local subsurface models from observed data and to generate waveforms for arbitrary source and receiver locations. Our model takes source and receiver geometries (and source parameters) as an input and generates the corresponding waveforms similar to physics-based numerical simulations, but without expensive computational cost or prior knowledge of the Earth's subsurface or earthquake sources. For example, we can generate (simulate) waveforms for future earthquakes, so that we can use them for seismic hazard analysis or for acquisition designing. More powerfully, the generated waveforms are considered as a different realization of the real observation, and they can be treated ``as if'' real observations, meaning that the usual processing can be applied to infer real earthquake processes and subsurface structures. Also, the generated waveforms at synthetic sensors can be considered as physically meaningful interpolation, and hence we can generate observed waveforms for virtual dense arrays. This can be used to identify previously un-identified phases due to spatial aliasing, as well as to detect earthquake events by increasing the stacking power over receivers. With these virtual sources and receivers, we can further apply advanced full wavefield-based imaging and inversion methods that typically require dense networks.  This capability sets apart our ML generation from traditional simulations, which are typically bounded to a specific representation of the earth. 
    We focus on a data-driven approach rather than physics-informed approaches to free ourselves from specific physics and pre-determined velocity and source models.

    To evaluate CGM-Wave, we use a dataset from the Geysers geothermal field, the world’s largest geothermal site with frequent induced seismicity due to energy production and water injection. Its high-density seismic network with abundant seismicity provides high-quality data, enabling rigorous model validation.
    We demonstrate the ability to generate synthetic wavefields at arbitrary sources and receivers that preserve detailed time-frequency-domain features and precise phases. 
    As a demonstration, we enhance the receiver density by 5 times more than the actual network along a line for a synthetic earthquake.
    This provides a spatially coherent earthquake gather that is easier to identify seismic phases.
    The frequency of the wavefields we generated is up to 15 Hz, and associated wavenumbers are much higher than the Nyquest wavenumber, according to the actual receiver spacing.
    By integrating observational data with data-driven representations of the Earth's physical processes, CGM-Wave establishes a transformative paradigm for wavefield generation, offering a robust tool for seismic imaging, hazard assessment, and beyond. 
    Additionally, CGM-Wave provides high-quality augmented datasets that can enhance many downstream ML applications, such as seismic phase picking, high-resolution seismic imaging, and ground motion predictions.

    We first present CGM-Wave, our diffusion-based conditional generative model. 
    Then, we evaluate CGM-Wave on the Geysers dataset with different source and receiver conditions. 
    Finally, we then discuss data augmentation, validation strategy, and limitation of CGM-Wave, followed by a brief conclusion.

\section{Methodology}
\label{sxn:methodology}

    CGM-Wave is built based on a well-defined probabilistic process via dual Markov chains that consist of forward and backward steps (Figure \ref{fig:fwork}):
    first, we perform a diffusion process that progressively transforms the seismic amplitude spectrum into a predefined noise distribution such as Gaussian noise;
    and second, we perform a denoising process that attempts to recover the original data through a network.
    This network includes a U-shaped transformer architecture to reconstruct the energy distribution and attenuation patterns for seismic amplitude spectrum synthesis. 
    To enable realistic generation of seismic wavefields,
    CGM-Wave has a module of PRM to estimate the corresponding phase spectrum. 

    \begin{figure*}[t!]
      \centering
      \makebox[\textwidth][c]{\includegraphics[width=1.0\textwidth]{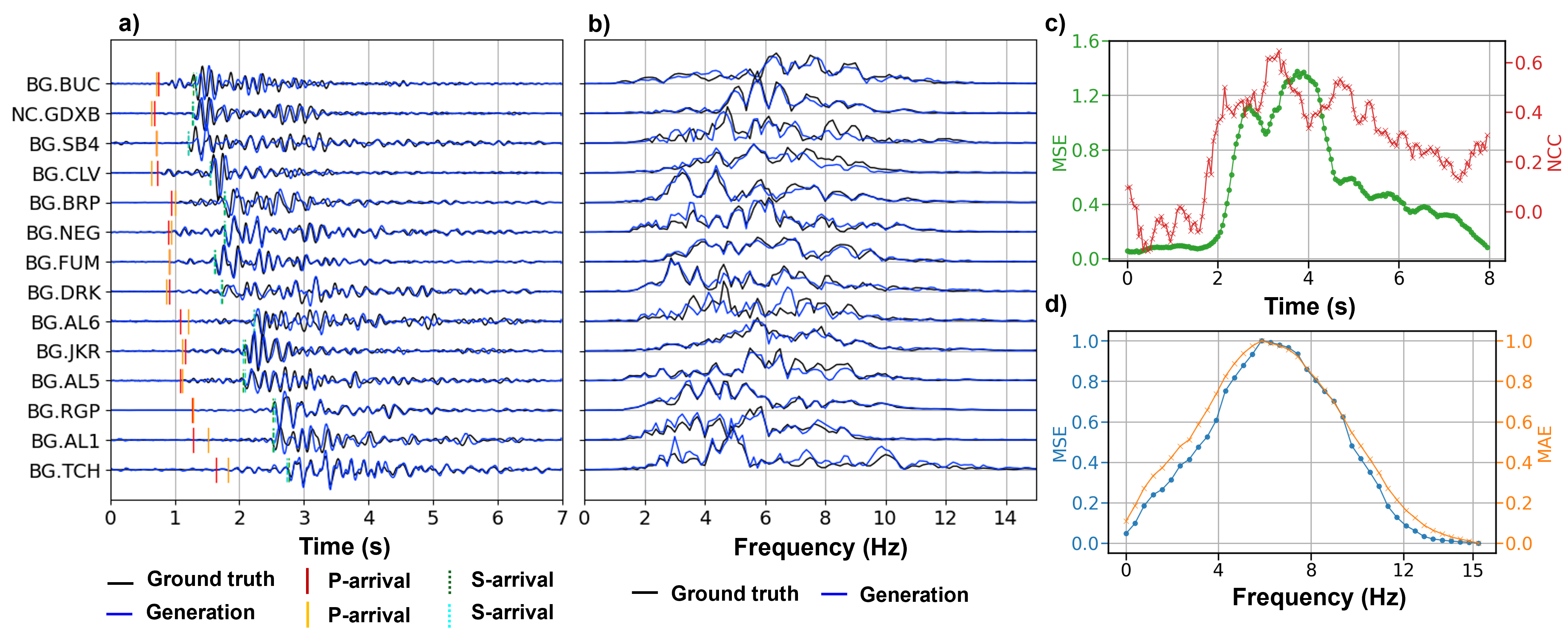}}
      \caption{
        Comparison of observed and synthetic waveforms on the validation dataset. Each trace is aligned by the distance from the epicenter.
        (a) Real data and synthetics generated by {CGM-Wave}, with vertical lines indicating P- and S-wave arrival times, calculated using PhaseNet. The IDs in each trace are the FDSN network and station codes.
        (b) Frequency spectra of observed and generated waveforms.
        Variations between generated waveforms and ground truth across different (c) epicentral distances and (d) frequency ranges.
      }
      \label{fig:fvalid}
    \end{figure*}

    \subsection{Diffusion Probabilistic Model}

    Given a data distribution ${q(x)}$ and data ${x_0}$ sampled from it, 
    the diffusion model provides a framework to model complexities of ${q(x)}$ by gradually corrupting data ${x_0}$. 
    In Step 1 (forward), Gaussian noise is incrementally introduced over ${K}$ steps, transforming ${x_0}$ into a sequence of latent variables ${(x_1,x_2,...,x_k,...,x_K)}$ through a Markov process. 
    Each transition is governed by a Gaussian diffusion kernel,
    \begin{equation}
        q(x_k | x_{k-1}) = \mathcal{N}(x_k;\sqrt{1-\beta_k}x_{k-1},\beta_k\mathbf{I}),
    \label{eq:shift}
    \end{equation}
    where $\beta_k \in (0,1)$ controls the magnitude of the noise at each step, and $\mathcal{N}(\mu,\sigma)$ represents a normal distribution with mean $\mu$ and standard deviation $\sigma$.
    Using the properties of the Gaussian kernel, the state $x_k$ can be expressed as a function of the initial data $x_0$ without stepwise computations of all intermediate states,
    \begin{equation}
        x_k = \sqrt{\bar{\alpha}_k}x_0+\sqrt{1-\bar{\alpha}_k}\epsilon,
    \label{eq:shift}
    \end{equation}
    in which ${\sqrt{\bar{\alpha}_k}=\prod_{i=1}^k(1-\beta_i)}$ represents the cumulative data retention factor, and ${\epsilon\sim\mathcal{N}(0,\mathbf{I})}$ denotes a Gaussian noise. 
    As ${k}$ approaches ${K}$, ${\bar{\alpha}_k}$ converges to ${0}$, and the forward process gradually transforms the data into pure Gaussian noise ${x_K\sim\mathcal{N}(0,\mathbf{I})}$, erasing all discernible structures. 
    In Step 2 (reverse), the reverse denoising process reconstructs the original data distribution by iteratively removing noise through a series of Markov transitions. 
    Starting from the standard Gaussian distribution ${p(x_k)\sim\mathcal{N}(0,\mathbf{I})}$,
    the process employs a learnable transition kernel ${p_\theta(x_{k-1} | x_k)}$ to progressively generate the final data distribution, ${p_\theta(x_0)}$.
    Each reverse transition is modeled as a Gaussian distribution,
    \begin{equation}
        p(x_{k-1} | x_k) = \mathcal{N}(x_{k-1};\mu_{\theta}(x_k,k),\sigma_{\theta}(x_k,k)\mathbf{I}),
    \label{eq:shift}
    \end{equation}
    where the mean ${\mu_\theta(\cdot)}$ and variance ${\sigma_\theta(\cdot)}$ are parameterized by the diffusion model ${\theta}$.
    Through this iterative denoising process, the network aims to approximate the true data distribution.

    \subsection{Conditional U-shaped Transformer}

    For high-dimensional data, diffusion models have demonstrated effectiveness when structured with a U-shaped architecture~\cite{ronneberger2015u}, a structure which naturally aligns with image-like inductive biases. 
    While this design excels at capturing continuous spatial features, our approach introduces a distinct inductive bias for seismic data, emphasizing the representation of energy distributions and attenuation patterns within the amplitude spectrum.
    The model is built around a 2-D transformer that extends its generative capabilities to a conditional distribution.
    Given a distribution ${p(x)}$, the model operates on noisy inputs ${x \sim p(x)}$ through a series of downsampling blocks that progressively extract multi-scale features.
    Each block (yellow box in Figure \ref{fig:fwork}) employs a combination of ResNet-based convolutions and cross-attention mechanisms to integrate both spatial and conditional information.
    A bottleneck further integrates features via ResNet layers, serving as a bridge between the encoder and decoder stages.
    The upsampling blocks reconstruct the data in reverse order, using \textit{skip connections} to incorporate intermediate features from corresponding downsampling stages.

    To enable guided synthesis conditioned on seismic attributes ${y}$, we employ the transformer as a denoising autoencoder that models the \emph{conditional} distribution ${p(x|y)}$.
    A learnable domain-specific encoder projects the various attributes $y$ into intermediate representations.
    These conditional representations are integrated into the model backbone through cross-attention mechanisms.
    The training process leverages pairs of amplitude spectra and corresponding attributes to optimize the model by minimizing the objective:
    \begin{equation}
        \mathcal{L}_\text{DM} = \mathbb{E}_{x,\epsilon\sim{\mathcal{N}(0,1)},y,t}\| \epsilon-\epsilon_\theta(x_t,y,t) \|_2^2 ,
    \label{eq:shift}
    \end{equation}
    where time step $t$ is uniformly sampled from a fixed Markov Chain with length ${T}$, $x_t$ denotes noisy data at diffusion time step ${t}$, and ${\epsilon_\theta(x_t,t)}$ represents the model's prediction of the noise ${\epsilon\sim{\mathcal{N}(0,1)}}$.

    \subsection{Phase Retrieval Method}

    Accurate phase reconstruction plays a crucial role in generating realistic wavefields, extending beyond just amplitudes in the frequency or time-frequency domain.
    This allows us to use the generated wavefields for phase identification and wave-equation-based applications.
    We develop a PRM that extends traditional Griffin-Lim algorithm~\cite{Griffin1984}, while generalizing it for broader applications.
    This method integrates iterative optimization and differentiable spectrogram transformations to reconstruct waveforms with minimal phase errors.
    Its framework includes forward and inverse spectrogram transformations, ensuring consistency between the time-domain signal and its frequency-domain representation.
    The forward pass starts with a pre-initialized phase spectrum, optimized using the AdamW optimizer (with a learning rate scheduler) to refine phase spectrum. 
    The complex spectrum reconstructed from the amplitude spectrum, and the updated phase spectrum is transformed back into the time domain. 
    The inverse pass applies the forward spectrogram transform to evaluate the reconstructed signal and refine alignment with the amplitude spectrum.
    
    The objective function is similar to the Griffin-Lim algorithm, and it combines amplitude spectrum consistency between the reconstructed ${A_i'(\omega)}$ and the generated ${A_i(\omega)}$, with sparsity-inducing $L_1$ and $L_2$ regularization applied to the waveform ${s_i(t)}$,
    \begin{equation}
        \mathcal{L}_\text{PR} = \frac{1}{N} \sum_{i=1}^{N} (\left\| |A_i(\omega)| - |A_i'(\omega)| \right\|^2+\lambda\|s_i(t)\|_1+\beta\|s_i(t)\|_2^2)  ,
    \label{eq:shift}
    \end{equation}
    where ${N}$ denotes the number of generations, and where $\lambda$ and $\beta$ represent the weighting factors that balance the loss components.
    Based on extensive empirical evaluations, both are set to $0.01$.
    The method optimizes multiple initializations of the phase spectrum from training dataset, selecting the result with the lowest loss across iterations to avoid local optima and further improve phase quality.
    The final optimized phase spectrum combines with the amplitude spectrum to generate high-fidelity seismic waveforms.  
    The comparison detailed in the~\ref{sec:s:phase} demonstrates that the PRM outperforms the traditional Griffin-Lim method, showing its robustness for phase retrieval from amplitude spectrum.

\section{Application}
\label{sxn:application}

    \begin{figure*}[t!]
        \makebox[\textwidth][c]{\includegraphics[width=0.8\textwidth]{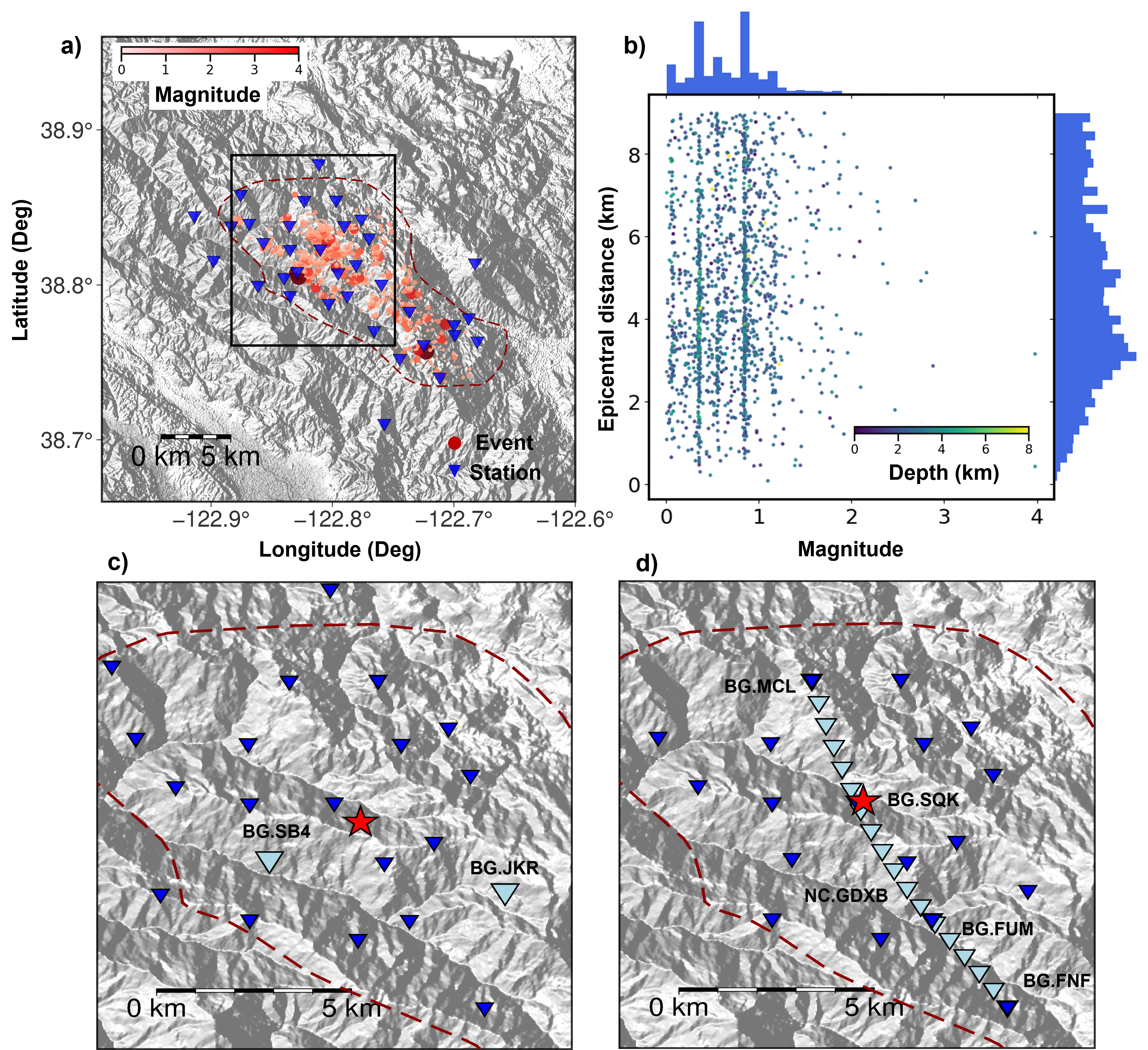}}
      \caption{
        Study region and dataset overview. 
        (a) Map of the Geysers geothermal field, with the dashed polygon delineating the reservoir. The background color illustrates the elevation, and the light color indicates high elevation. 
        Blue triangles indicate stations, while circles mark earthquake events used in this study. The color of the circles and size scale with magnitude.
        (b) Distributions of event magnitude and source-to-station distance for the earthquakes used in this study for training and validation datasets. 
        (c) Map of a validation seismic event (red star) and the stations (dark blue), along with a synthetic linear array (light blue) across four real stations.
        (d) Spatial layout of the included and excluded stations associated within the validation event.
      }
      \label{fig:fmap}
    \end{figure*}

    The Geysers geothermal field (Figure~\ref{fig:fmap}a) in Northern California stands as one of the globe's largest and most enduring sources of geothermal energy since the early 1960s~\cite{khan2010geysers}. 
    The extensive exploration and use of this field have led to significant hydrothermal activity and a high rate of earthquakes. 
    We use the Geysers geothermal field as an example site to demonstrate CGM-Wave. 
    
    \subsection{Dataset Preparation and Training}
    
    The dataset comprises raw waveforms recorded by $35$ stations from the Berkeley Geysers and Northern California networks between $2021$ and $2022$. 
    We extract $500$ seismic events (Figure~\ref{fig:fmap}a) with arrivals recorded across the networks, with $50$ of them reserved for validation and the remaining used for training.  This results in a total of 17,500 earthquake records, of which 1,750 belong to the validation set. 
    The datasets focus on vertical-component waveforms from  small-magnitude earthquakes (${M_L<2}$) occurring within the reservoir region, less than 9 km epicentral distances (Figure~\ref{fig:fmap}b).
    To preprocess the raw data, we remove the mean and trend, followed by bandpass filtering between 0.01–15 Hz and resampling to $100$ Hz.
    The amplitude spectra are normalized globally to stabilize the training process of the model by ensuring numerical consistency across features while preserving the amplitude balances between receivers and earthquakes.
    The normalization parameters were recorded to enable restoration of the amplitude spectra after prediction.
    After predictions, normalized amplitude spectra are generated, and we apply denormalization restored using the normalization parameters.
    To enable the model to capture patterns of waveform propagation across the stations with reasonable spatial continuity, we employ a physical-based data augmentation strategy using dynamic time warping to expand our dataset (\ref{sec:s:aug}).
    This method expands the dataset to expose the model to a wider range of source-station spatial configurations.
 
    \begin{figure*}[t!]
      \centering
      \makebox[\textwidth][c]{\includegraphics[width=1.0\textwidth]{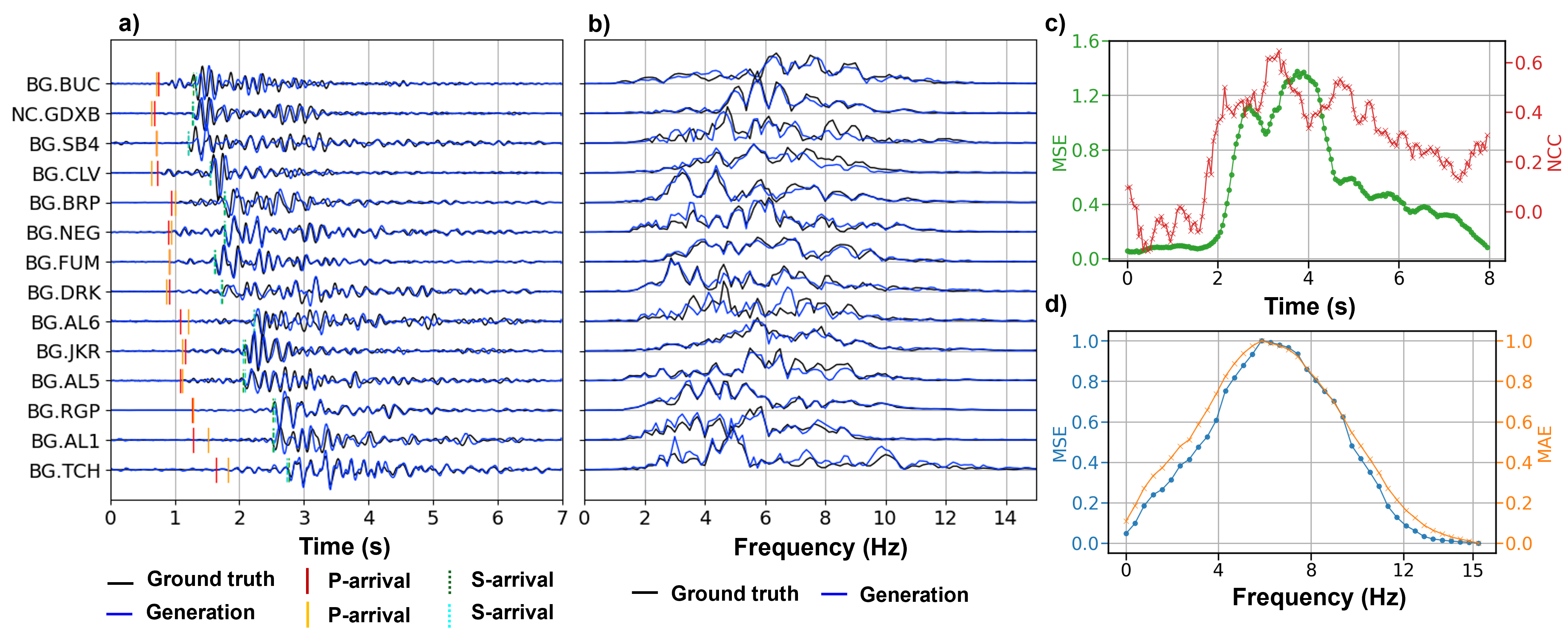}}
      \caption{
        Comparison of observed and synthetic waveforms on the validation dataset. Each trace is aligned by the distance from the epicenter.
        (a) Real data and synthetics generated by {CGM-Wave}, with vertical lines indicating P- and S-wave arrival times, calculated using PhaseNet. The IDs in each trace are the FDSN network and station codes.
        (b) Frequency spectra of observed and generated waveforms.
        Variations between generated waveforms and ground truth across different (c) epicentral distances and (d) frequency ranges.
      }
      \label{fig:fvalid}
    \end{figure*}

    To train the diffusion model, we adopt a forward diffusion process that integrates noise scheduling and conditioning inputs. 
    During training, Gaussian noise is incrementally added to the amplitude spectra ${x}$ at randomly sampled timesteps (${t \in T}$). 
    This controlled addition of noise ensures a smooth transition from the original data to a noise-dominated state by simulating the forward diffusion dynamics.
    The model is optimized to predict the noise residual for denoising.
    The amplitude spectra computed by STFT serve as input data ${x}$.
    The generated amplitude spectra are conditioned on descriptive embeddings derived from input attributes, including earthquake magnitude, depth, source coordinates, and station locations.
    Optimization is performed using the AdamW optimizer with an initial learning rate of ${1 \times 10^{-5}}$, complemented by a learning rate scheduler to ensure smooth convergence.
    
    \subsection{Validation of Model Performance}
    \subsubsection{Wavefield generation within validation dataset}

    We evaluate the generated waveforms against the ground truth from a seismic event randomly extracted from the validation dataset. Hence the network has not encountered the validation earthquake during training. We test its ability to position a source at any location within the area of interest and assign it a magnitude of our choice. Note that in this section, we use the stations at physical locations; while in the next sections, we put stations at arbitrary locations as well.
    As shown in Figure~\ref{fig:fvalid}a, the generated waveforms not only closely match the real data, but they also demonstrate consistent P-wave and S-wave arrivals detected using PhaseNet~\cite{zhu2019phasenet}, despite the absence of phase information as conditioning inputs.
    
    The generated waveforms replicate the amplitude decay patterns of real data over time, indicating the efficacy of incorporating source and station locations into the conditioning inputs.
    Furthermore, the frequency (Figure~\ref{fig:fvalid}b) and time-frequency attributes (Figure~\ref{fig:fcompare}) of the generated results exhibit resemblance to those of the real data.    
    To quantitatively evaluate the model performance, we compute Mean Squared Error (MSE), Mean Absolute Error (MAE), and Normalized Cross-Correlation (NCC) between ground truth and generations over all 50 earthquakes in the validation set in the time and frequency domains (Figures~\ref{fig:fvalid}c and \ref{fig:fvalid}d).
    The primary discrepancies are observed within the $0–3$ s time window and $4–8$ Hz frequency band, likely due to the complexity of transient signals and mid-range frequencies.
    Additionally, we compare our model against those produced by other deep learning-based wavefield generation approaches. 
    The results demonstrate that CGM-Wave achieves superior quality in generating amplitude spectra, as detailed in~\ref{sec:s:compare}.

    \begin{figure*}[t!]
      \centering
      \makebox[\textwidth][c]{\includegraphics[width=0.8\textwidth]{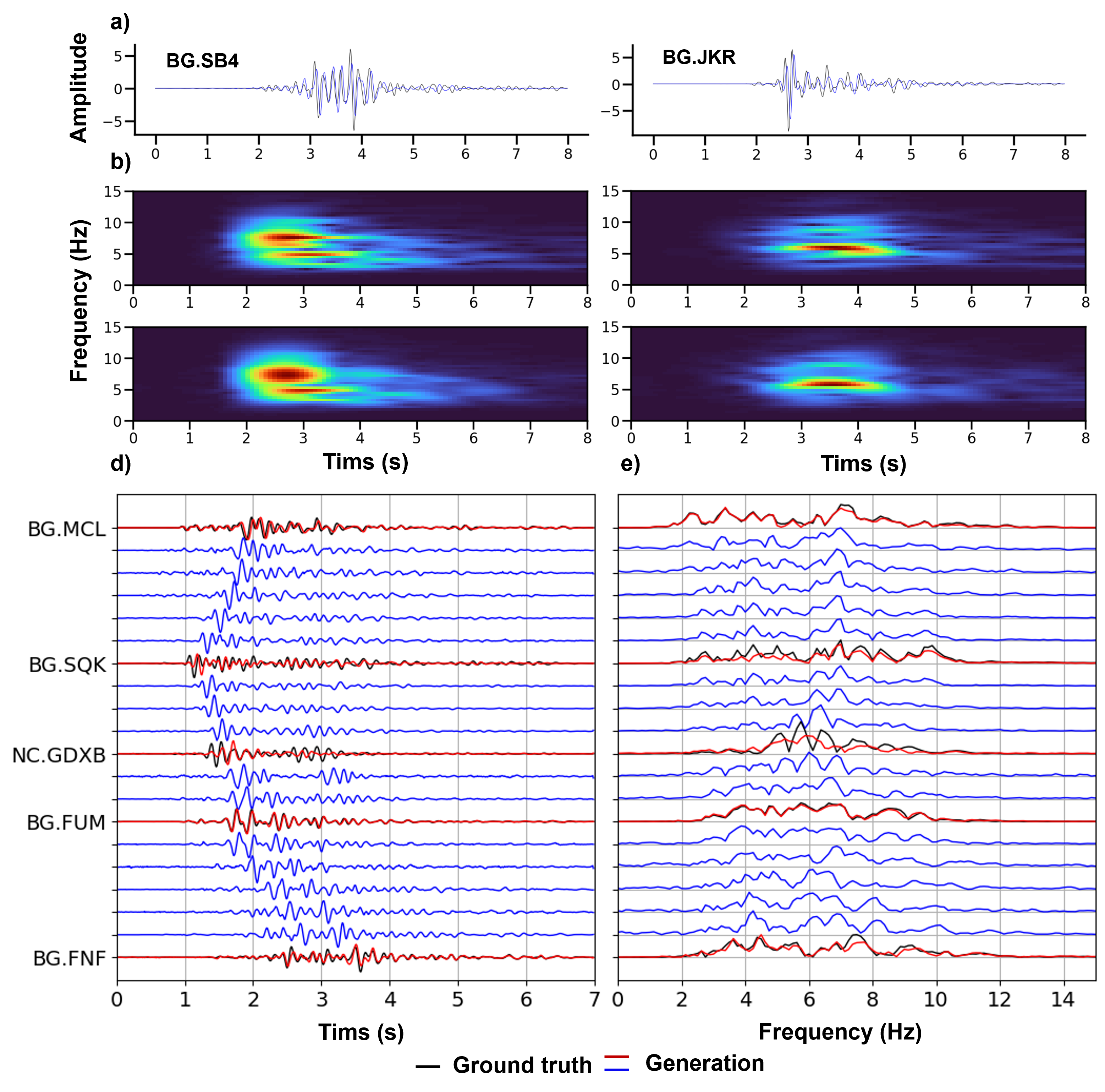}}
      \caption{
        (a) Synthetic seismic wavefields for two excluded stations (first row) from training, with (b) ground truth spectrograms (first row) and generated spectrograms (second row) for comparison.
        (d) Frequency-domain and (e) time-domain comparisons of observed and synthetic wavefields along the linear array in real (red) and synthetic (blue) stations.
      }
      \label{fig:farray}
    \end{figure*}

    \subsubsection{Wavefield generation with arbitrary sources and arbitrary receivers}

    In Figure \ref{fig:fvalid}, we validate synthetic wavefields by comparing the wavefields for an earthquake in the validation dataset. 
    For the purpose of generating wavefields at arbitrary sources and receivers, in this subsection, we further validate generated wavefields at receivers which are excluded from training, and earthquakes which are in the validation dataset.
    
    We intentionally exclude two stations from training datasets (light blue triangles in Figure~\ref{fig:fmap}c), and then we task the model with generating wavefields at these excluded stations.
    Comparisons of the generated amplitude spectra and reconstructed waveforms with ground truth data reveal an alignment (Figures~\ref{fig:farray}a and~\ref{fig:farray}b).
    The PRM minimizes phase inconsistencies, while preserving waveform sparsity and smoothness, ensuring the physical realism of the generated wavefields. 
    These wavefields replicate key spectral and temporal features observed in the ground truth data, including phase arrivals and frequency-dependent amplitude variations. 
    A small time shift is observed in the generated wavefields, consistent with some traces shown in Figure \ref{fig:fvalid}a. 
    This shift likely arises from a bulk phase adjustment across all frequencies during the PRM step. 
    Although future tests are needed to investigate such phase shift, the time shift remains minimal, and the generated wavefields retain their utility for most of downstream applications.
    This test emphasizes the capability of the model to encode data-driven representations of Earth's physical processes, enabling the data generation in regions with sparse observational coverage, without requiring prior knowledge of subsurface properties or seismic sources. 

    \subsubsection{Generation of earthquake records with a linear receiver array}

    As a potential application of {CGM-Wave}, we simulate seismic waveforms in a synthetic line array that spans four real stations from a validation seismic event (Figure~\ref{fig:fmap}d).
    This setup enables the generation of seismic waveforms both at the stations and in the interpolated regions between them.
    Shot gathers have been used for structural imaging. 
    They can display the effects of earthquake rupture as well as measure the amplitude distribution and peak frequencies along the dense sampling.
    Hence, the generation of shot gathers is an important tasks for {CGM-Wave} for further applications.
    
    The generated wavefields closely match the observations at the four stations with known data (Figures~\ref{fig:farray}d).
    The generated amplitude spectrum reveals alignment with the expected spectral characteristics, indicating that the model captures the frequency content of seismic signals (Figure~\ref{fig:farray}e).
    For the virtual stations between the real stations, the generated waveforms exhibit a physically consistent move-out pattern of phase and amplitudes. 
    Based on the velocity models presented by \cite{Nayak2018,Lin2018}, the velocities on the north side are smaller than the south side especially south of NC.GDXB.
    This is consistent with what we observe in the generated wavefields.
    The south side contains a slow velocity anomaly (separation of upper and lower reservoir; \cite{Gritto2022}), which can be related to the phase nearly 1 second after the direct S wave. The spatial continuity of this second phase becomes clearer after the wavefield generation.
    The velocity anomaly starts around NC.GDXB, which potentially causes the complicated wavefields with later phases in this and two south synthetic receivers.
    Further investigation is needed, but the generated wavefields follow the general velocity structures obtained by seismic tomography.
    Such consistency insists its potential to generalize beyond the training data, and it suggests that the model effectively encodes the wave propagation dynamics.

\section{Discussion}
\label{sxn:discussion}

    The wavefields generated by CGM-Wave offer a novel approach for improving results of various seismic applications.
    For example, the generated waveforms can enhance ground motion prediction models by providing generative observations for regions where physical data collection is sparse.
    Moreover, by synthesizing accurate wavefields in areas where available sources and receivers are limited, we enhance the spatial resolution of seismic imaging.
    This capability supports the detailed modeling of subsurface structures and more comprehensive seismic analyses.

    However, the absence of explicit physical constraints, such as governing equations for wave propagation and earthquake rupture models, limits CGM-Wave to ensure the physical consistency of the generated wavefields.
    Such physical constraints can also enhance model's physical interpretability.
    Observed discrepancies between synthetic and real wavefields highlight the combined influence over source-path-site effects.
    Although data augmentation has been employed to partially mitigate this issue by improving data diversity, it also risks introducing biases rooted in the training, constraining the model’s generalization to unseen scenarios.     
    While the generated wavefields to densify the existing network are encouraging, stably generating wavefields at arbitrary station locations remains challenging, when considering local structural anomalies especially for high-frequency wave generation. 
    For more robust generation, we need to expand the conditional variable space to include source attributes, such as moment tensors, stress drop distributions, and rupture directionality, which are currently not considered and included in the error of generations.

    Interestingly, based on our experiments, generating wavefields for arbitrary earthquakes is relatively easy compared to generating wavefields for arbitrary stations. 
    We speculate that a possible reason is that the earthquakes distribute much denser than the receivers, although the earthquakes distribution is in 3-D, and the variation of the site effects among receivers are larger than the variation of the structure around sources. 
    Needless to say, the density of both sources and receivers is essential for the model’s performance. 
    Sparse data coverage limits the model's ability to capture fine-grained spatial and temporal variations.

    Finally, training diffusion models presents another challenge due to their iterative nature, requiring meticulous hyper-parameter tuning.
    Exploring adaptive noise schedules and model compression techniques could enhance training efficiency while maintaining performance. 
    These advancements will be critical in making diffusion-based approaches more scalable and applicable to larger and more diverse datasets.

\section{Conclusions}
\label{sxn:conclusions}

    This study presented CGM-Wave, demonstrating the potential of diffusion models for generating high-fidelity seismic waveforms. 
    By integrating the STFT and an adaptive phase retrieval, our CGM-Wave framework produces wavefields that accurately capture temporal, spectral, and noise characteristics of real seismic data.
    Validation using data from the Geysers geothermal field confirms the method's ability to replicate seismic wavefields, including reasonable P- and S-wave arrivals, waveform moveout related to the source-receiver offsets, and amplitude decay in time and space domains.
    Quantitative evaluation of the generated waveforms indicates alignment with real data, achieving spectral coherence across the target frequency range. 
    
    Our results for generated wavefields for both phase and amplitude highlight the applicability of our model to enhancing seismic imaging, phase picking, magnitude estimation, and source characterization in data-scarce regions.
    Future efforts will focus on integrating physical constraints to further improve realism for seismic hazard assessment and subsurface characterization.

\clearpage
\appendix
\section{Training Dataset Augmentation}
\label{sec:s:aug}

    \begin{figure}[t!]
      \centering
      \includegraphics[width=0.8\textwidth]{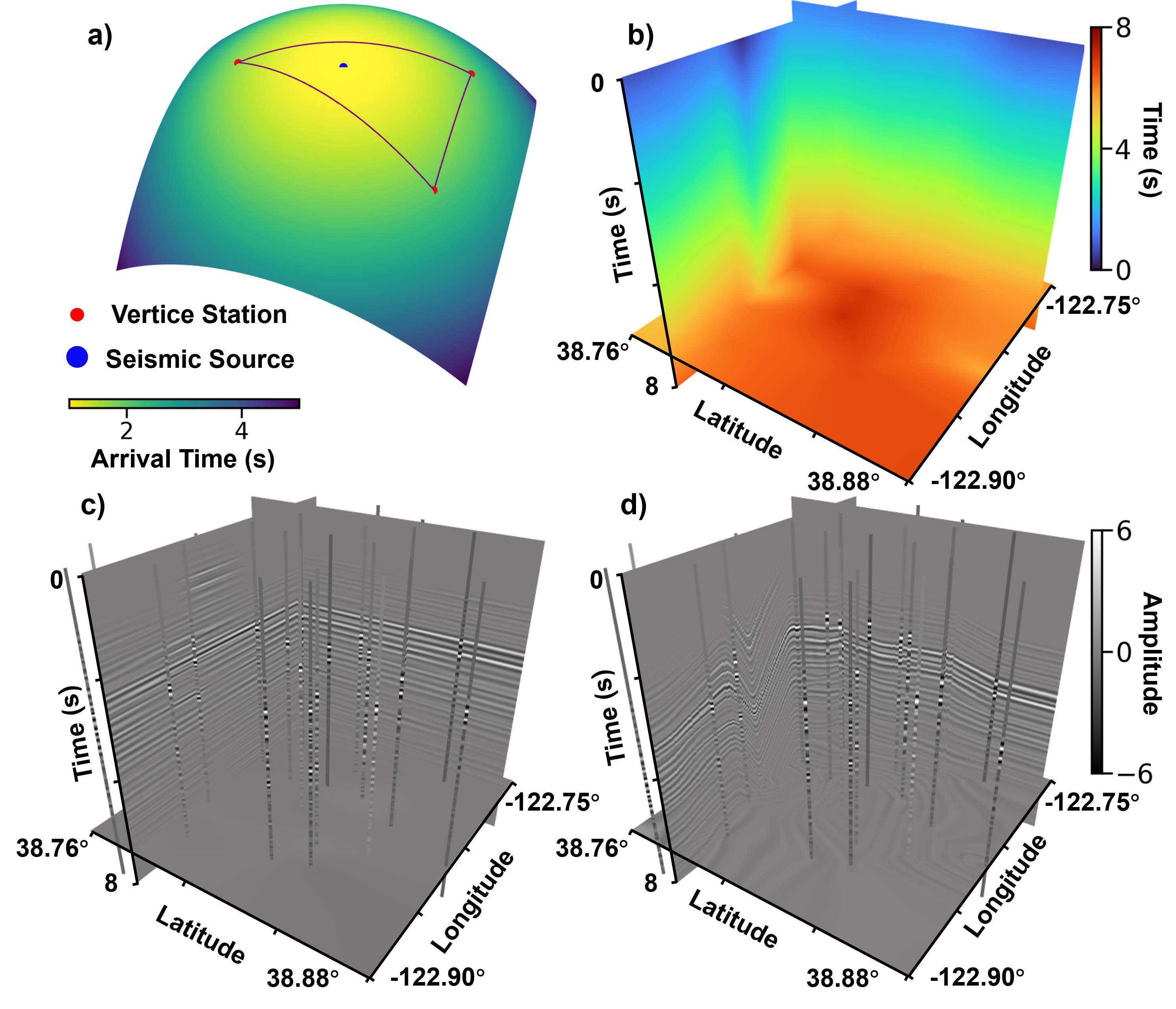}
      \caption{
        Data augmentation for expanding the training dataset.
        (a) Phase arrival times estimated from surrounding stations at epicenter using hyperbolic equations based on constant velocity.
        (b) Shifts computed from DTW and interpolation.
        (c) Alignment of waveforms across all stations using dynamic time warping (DTW), followed by interpolation in the flattened space;
        vertical lines represent the horizontally aligned time-series.
        (d) Wavefield produced by restoring the interpolated aligned volume;
      }
      \label{fig:faug1}
    \end{figure}

    \begin{figure*}[t!]
      \centering
      \includegraphics[width=1.0\textwidth]{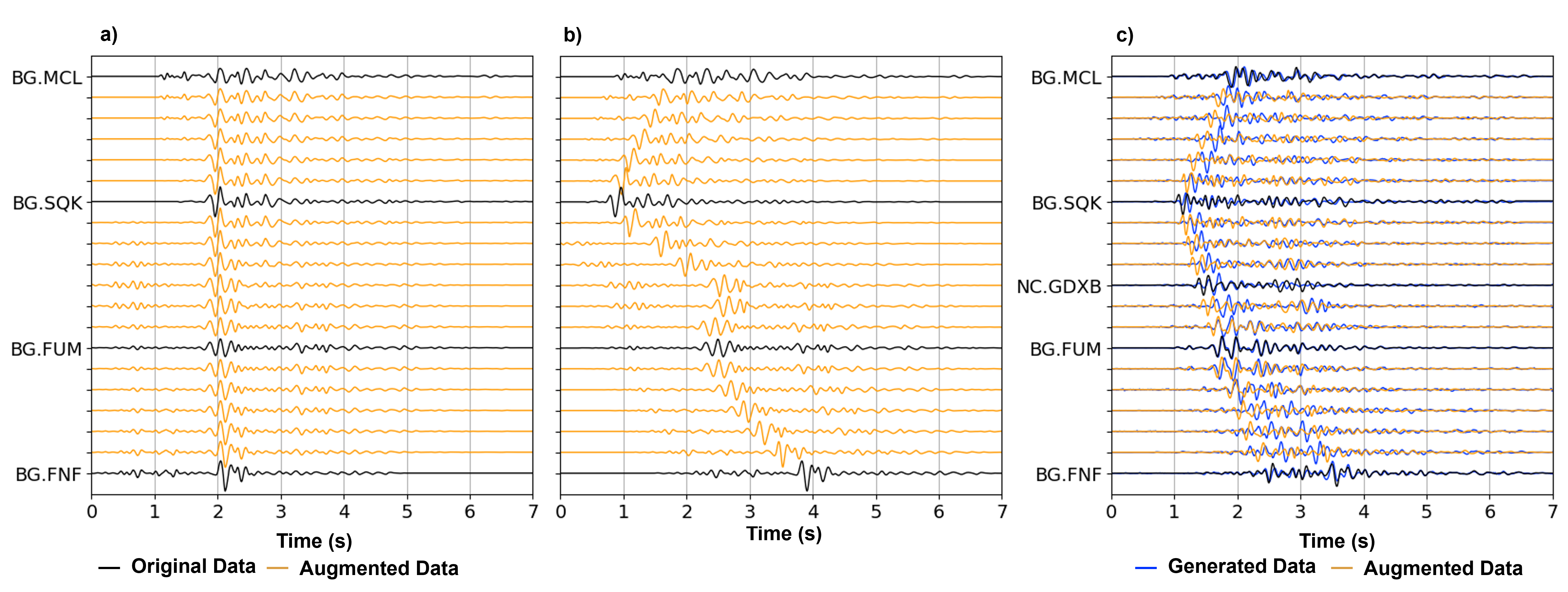}
      \caption{
        Comparison of real and augmented seismic waveforms.
        (a) Aligned and (b) restored seismic wavefields along a linear array for a seismic event shown in Figure~\ref{fig:farray}d. 
        (c) Comparison between the augmented (orange) and generated (blue) wavefield.
      }
      \label{fig:faug2}
    \end{figure*}

    Seismic monitoring often suffers from sparse station coverage due to the high costs and logistical challenges of seismic network deployment. 
    While the diffusion model demonstrates the ability to generate synthetic seismic wavefield with realistic temporal and spectral characteristics at and near station locations, uncertainties become large in regions lacking dense observational data. 
    The wavefield generated for these unmonitored regions may not fully adhere to wave propagation dynamics in terms of phase characteristics and amplitude decay, as constrained by physical principles like the wave equations.
    To mitigate the effects of sparse data distributions, this study introduces a physical-based data augmentation strategy to expose the model to diverse spatial configuration scenarios.
    Specifically, we introduce a phase-guided interpolation method to generate synthetic data for interstation regions using known waveforms from the training dataset. 
    
    Using the surrounding stations as vertices, we first perform triangulation to partition the study area and identify the triangle enclosing seismic source. 
    To interpolate the seismic data at the epicenter, P-wave arrivals are detected at the vertex stations within the triangle using a phase-picking approach. 
    Assuming a constant velocity model and direct wave propagation, the P-wave arrival time at the epicenter is estimated via a hyperbolic equation (Figure~\ref{fig:faug1}a).
    Then we employ dynamic time warping (DTW) to align waveforms recorded at the vertex stations. 
    The aligned waveforms are used to interpolate the waveform at the epicenter. 
    Residual arrival times between the epicenter and the vertex stations are used to restore this interpolated waveform to its original phase characteristics.
    Following the extrapolation of the epicentral waveform, DTW is further applied to align all waveforms across the area, and to yield the associated temporal shifts. 
    Interpolation is conducted within this aligned space to generate a shift volume (Figure~\ref{fig:faug1}b) and produce a wavefield with spatially coherent phases (Figure~\ref{fig:faug1}c and Figure~\ref{fig:faug2}a). 
    Finally, the shift volume is employed to restore the aligned wavefield to its original configuration (Figure~\ref{fig:faug1}d and Figure~\ref{fig:faug2}b). 
    This workflow enriches the spatial coverage of dataset by synthesizing waveforms for unmonitored~regions.

    To ensure that the model simulates wavefield governed by learned physical principles rather than mimicking the behavior introduced from data augmentation, we compared the synthetic wavefields with those produced through the augmentation process (Figure~\ref{fig:faug2}c). 
    While the augmented waveforms rely on assumptions of constant velocity and linear interpolation, our model synthesizes waveforms that account for more complex spatial and temporal dependencies, reflecting the variability in real data.
    The analysis indicates that the model is not simply reproducing the augmented data but instead generalizes beyond its limitations by encoding underlying wave propagation dynamics.
    Such augmentation aims to expose the model to a wider range of source-station geometries, improving its ability to generate physically plausible waveforms across diverse spatial configurations.
    However, note that augmentation method may introduce biases, such as the gradual transitions in seismic phase among stations, potentially limiting model generalization to new geological settings. 
    Although the data-drive approaches such as using DTW proposed here make our method easy to apply to any other datasets, incorporating synthetic datasets derived from physical simulations using accurate subsurface models may address these biases and further enhance the training data diversity.

\section{Comparative Experiment with Other Phase Retrieval Methods}
\label{sec:s:phase}

    \begin{figure*}[t!]
      \centering
      \includegraphics[width=0.8\textwidth]{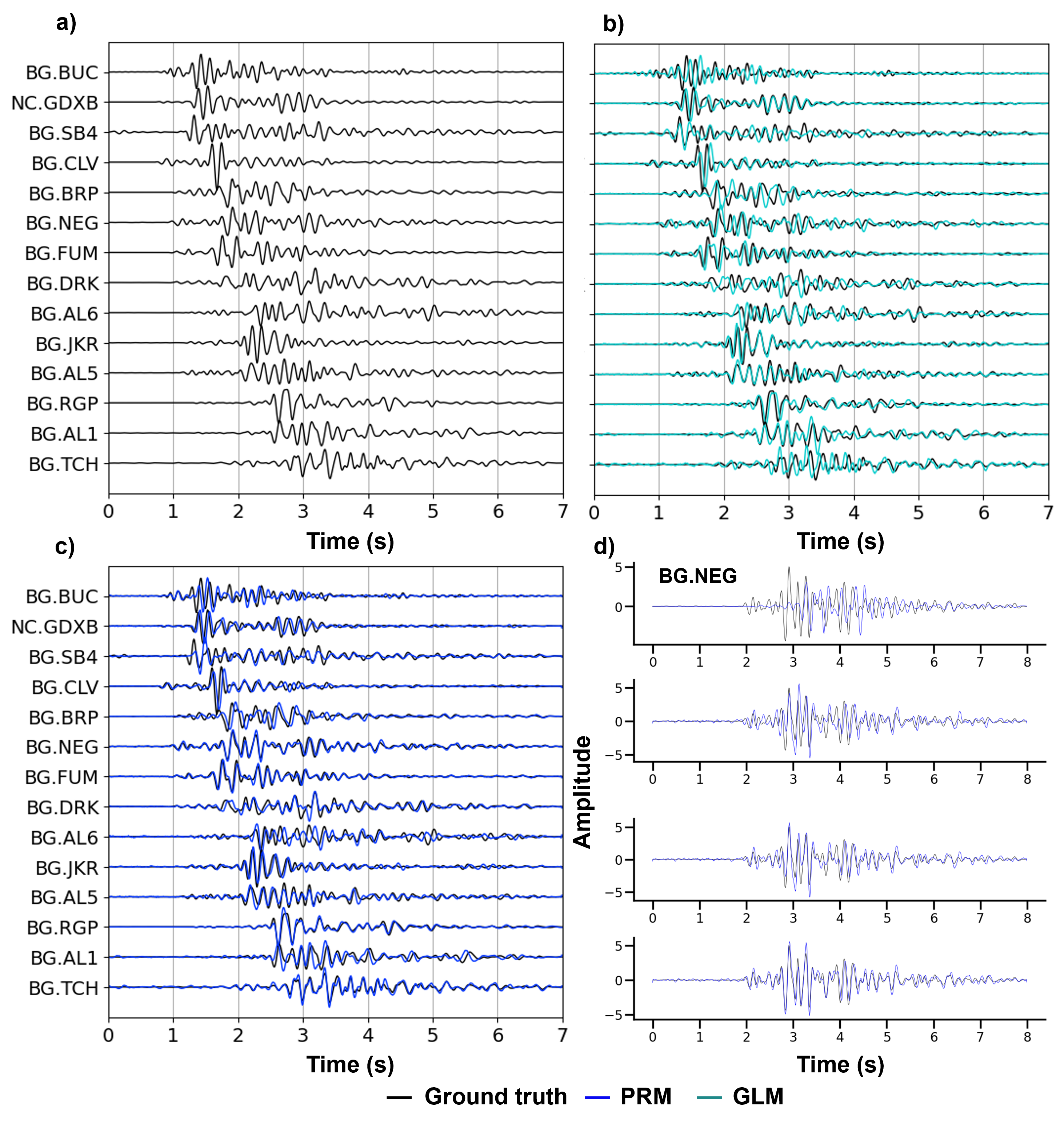}
      \caption{
        Comparison of different phase retrieval methods.
        (a) Ground-truth seismic wavefield (black).
        (b) Comparison between the ground-truth and the reconstructed wavefields (cyan) using the Griffin-Lim method.
        (c) Comparison between the ground-truth and the reconstructed wavefield (blue) using the  PRM method.
        (d) Iterative updates of the phase spectrum within the PRM method, showing the  refinement of the reconstructed wavefields (blue) towards the ground-truth (black).
      }
      \label{fig:fphase}
    \end{figure*}

  To assess the performance of the proposed Phase Retrieval Method (PRM), we compare seismic wavefields reconstructed using the PRM and the conventional Griffin-Lim algorithm (GL) with the ground truth (Figure~\ref{fig:fphase}a). 
  While the GL algorithm can approximately recover phase characteristics, noticeable mismatches with the ground-truth wavefield persist (Figure~\ref{fig:fphase}b). 
  In contrast, the PRM effectively mitigates these discrepancies (Figure~\ref{fig:fphase}c), delivering improved quality of phase retrieval and achieving reconstructions that more closely align with the ground truth. 
  As shown in Figure~\ref{fig:fphase}d, the iterative updates of the phase spectrum within this method demonstrate the progressive refinement of the wavefield toward the desired result.
    
\section{Comparative Experiment with Other Deep Learning Models}
\label{sec:s:compare}

    \begin{figure*}[t!]
      \centering
      \includegraphics[width=0.8\textwidth]{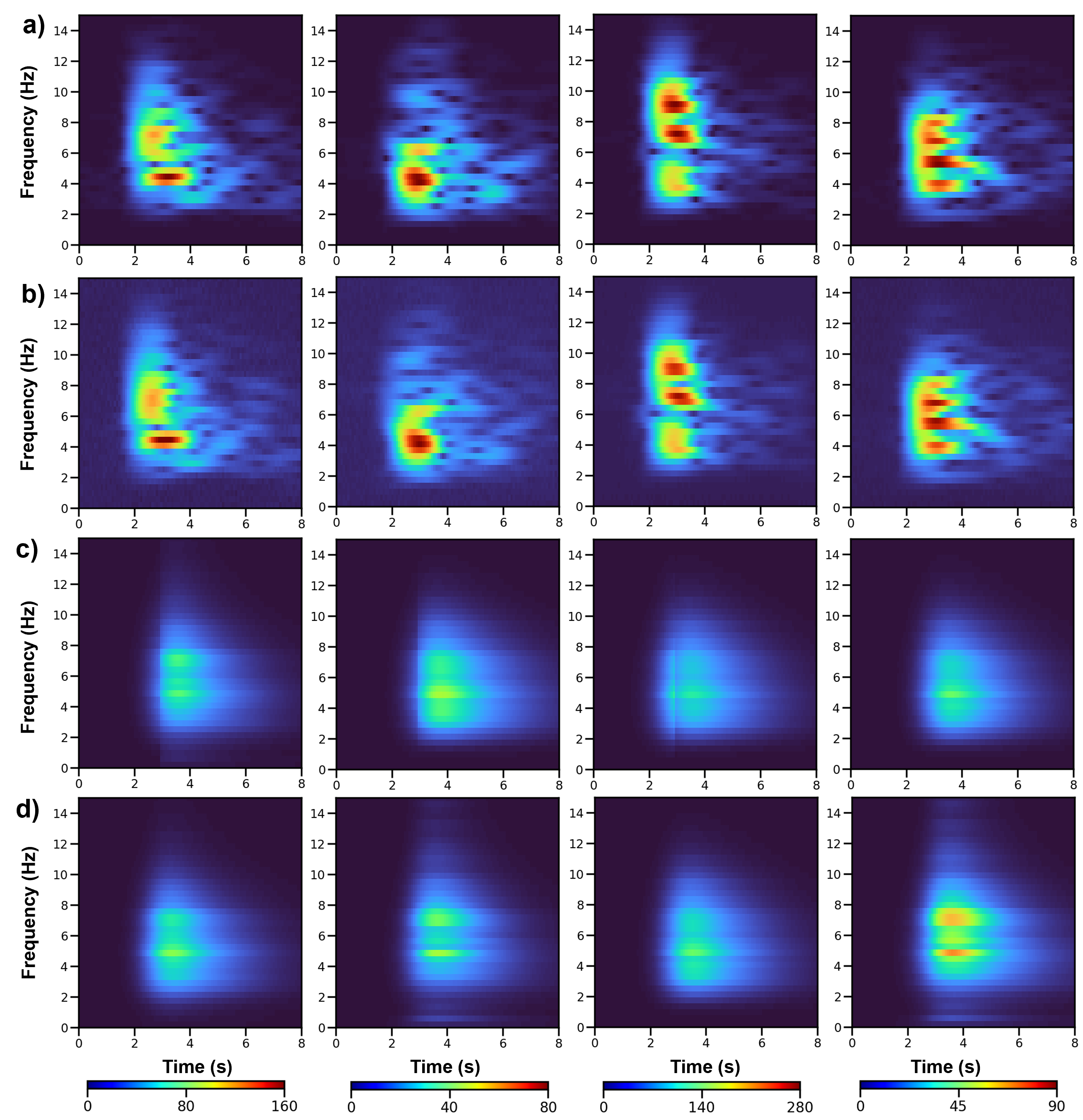}
      \caption{
        Comparison of real and synthetic spectrograms across ML models on the validation dataset.
        (a) Ground truth spectrograms.
        (b) The diffusion model results.
        (c) The CGM-GM (VAE) results.
        (d) The VAE-GAN model results.
      }
      \label{fig:fcompare}
    \end{figure*}

    To evaluate the performance of the proposed diffusion model, we conduct a comparison experiment using the validation dataset to benchmark the generated spectrograms against ground truth (Figure~\ref{fig:fcompare}a) and other state-of-the-art models, including a Variational Autoencoder (VAE; CGM-GM, \cite{ren2024learning}) and a VAE-GAN model.
    The diffusion model (Figure~\ref{fig:fcompare}b) produces spectrograms that closely align with the ground truth, preserving spectral continuity and faithfully reconstructing patterns of amplitude decay and noise characteristics. 
    In contrast, the VAE model (Figure~\ref{fig:fcompare}c) captures the overall spectral structure but exhibits excessive smoothing, particularly in high-frequency regions, which diminishes the fidelity of fine-grained details. 
    The VAE-GAN model (Figure~\ref{fig:fcompare}d) improves spectral sharpness and contrast, particularly in higher-frequency bands, but introduces artifacts and lacks the phase fidelity achieved by the diffusion model.
    This limitation likely stems from the adversarial training dynamics prioritizing visual realism over physical accuracy. 
    The results demonstrate that the diffusion model outperforms the VAE and VAE-GAN models by generating spectrograms that more accurately replicate the spectral and temporal complexities of real data.

\section*{Acknowledgments}
This work was supported by the Laboratory Directed Research and Development Program of Lawrence Berkeley National Laboratory (LBNL) under U.S. Department of Energy Contract No. DE-AC02-05CH11231,  the U.S. Department of Energy, Office of Energy Efficiency and Renewable Energy (EERE), Geothermal Technologies Office, under Award Number DE-AC02–05CH11231 with LBNL, and Utah FORGE R\&D project under Award Number 6-3656 with LBNL.

\bibliographystyle{unsrt}  
\bibliography{main}

\begin{thebibliography}{10}

\bibitem{2019booknakata}
Nori Nakata, Lucia Gualtieri, and Andreas Fichtner.
\newblock {\em {Seismic Ambient Noise}}.
\newblock Cambridge Univ. Press, 2019.

\bibitem{mousavi2020machine}
S~Mostafa Mousavi and Gregory~C Beroza.
\newblock A machine-learning approach for earthquake magnitude estimation.
\newblock {\em Geophysical Research Letters}, 47(1):e2019GL085976, 2020.

\bibitem{mousavi2022deep}
S~Mostafa Mousavi and Gregory~C Beroza.
\newblock Deep-learning seismology.
\newblock {\em Science}, 377(6607):eabm4470, 2022.

\bibitem{ni2024wavefield}
Yiyu Ni, Marine~A Denolle, Qibin Shi, Bradley~P Lipovsky, Shaowu Pan, and J~Nathan Kutz.
\newblock Wavefield reconstruction of distributed acoustic sensing: Lossy compression, wavefield separation, and edge computing.
\newblock {\em Journal of Geophysical Research: Machine Learning and Computation}, 1(3):e2024JH000247, 2024.

\bibitem{liu2024generative}
Qi~Liu and Jianwei Ma.
\newblock Generative interpolation via a diffusion probabilistic model.
\newblock {\em Geophysics}, 89(1):V65--V85, 2024.

\bibitem{wang2021seismogen}
Tiantong Wang, Daniel Trugman, and Youzuo Lin.
\newblock Seismogen: Seismic waveform synthesis using gan with application to seismic data augmentation.
\newblock {\em Journal of Geophysical Research: Solid Earth}, 126(4):e2020JB020077, 2021.

\bibitem{esfahani2023tfcgan}
Reza~DD Esfahani, Fabrice Cotton, Matthias Ohrnberger, and Frank Scherbaum.
\newblock Tfcgan: Nonstationary ground-motion simulation in the time--frequency domain using conditional generative adversarial network (cgan) and phase retrieval methods.
\newblock {\em Bulletin of the Seismological Society of America}, 113(1):453--467, 2023.

\bibitem{chen2024deep}
Guoyi Chen, Junlun Li, and Hao Guo.
\newblock Deep generative model conditioned by phase picks for synthesizing labeled seismic waveforms with limited data.
\newblock {\em IEEE Transactions on Geoscience and Remote Sensing}, 2024.

\bibitem{li2020seismic}
Kunhong Li, Song Chen, and Guangmin Hu.
\newblock Seismic labeled data expansion using variational autoencoders.
\newblock {\em Artificial Intelligence in Geosciences}, 1:24--30, 2020.

\bibitem{ren2024learning}
Pu~Ren, Rie Nakata, Maxime Lacour, Ilan Naiman, Nori Nakata, Jialin Song, Zhengfa Bi, Osman~Asif Malik, Dmitriy Morozov, Omri Azencot, et~al.
\newblock Learning physics for unveiling hidden earthquake ground motions via conditional generative modeling.
\newblock {\em arXiv preprint arXiv:2407.15089}, 2024.

\bibitem{gao2020zero}
Rui Gao, Xingsong Hou, Jie Qin, Jiaxin Chen, Li~Liu, Fan Zhu, Zhao Zhang, and Ling Shao.
\newblock Zero-vae-gan: Generating unseen features for generalized and transductive zero-shot learning.
\newblock {\em IEEE Transactions on Image Processing}, 29:3665--3680, 2020.

\bibitem{niu2020lstm}
Zijian Niu, Ke~Yu, and Xiaofei Wu.
\newblock Lstm-based vae-gan for time-series anomaly detection.
\newblock {\em Sensors}, 20(13):3738, 2020.

\bibitem{naiman2024iclr}
Ilan Naiman, N.~Benjamin Erichson, Pu~Ren, Michael~W. Mahoney, and Omri Azencot.
\newblock Generative modeling of regular and irregular time series data via koopman vaes.
\newblock In {\em {ICLR}}. OpenReview.net, 2024.

\bibitem{yang2023rapid}
Yan Yang, Angela~F Gao, Kamyar Azizzadenesheli, Robert~W Clayton, and Zachary~E Ross.
\newblock Rapid seismic waveform modeling and inversion with neural operators.
\newblock {\em IEEE Transactions on Geoscience and Remote Sensing}, 61:1--12, 2023.

\bibitem{kingma2021variational}
Diederik Kingma, Tim Salimans, Ben Poole, and Jonathan Ho.
\newblock Variational diffusion models.
\newblock {\em Advances in neural information processing systems}, 34:21696--21707, 2021.

\bibitem{rombach2022high}
Robin Rombach, Andreas Blattmann, Dominik Lorenz, Patrick Esser, and Bj{\"o}rn Ommer.
\newblock High-resolution image synthesis with latent diffusion models.
\newblock In {\em Proceedings of the IEEE/CVF conference on computer vision and pattern recognition}, pages 10684--10695, 2022.

\bibitem{ho2022cascaded}
Jonathan Ho, Chitwan Saharia, William Chan, David~J Fleet, Mohammad Norouzi, and Tim Salimans.
\newblock Cascaded diffusion models for high fidelity image generation.
\newblock {\em Journal of Machine Learning Research}, 23(47):1--33, 2022.

\bibitem{croitoru2023diffusion}
Florinel-Alin Croitoru, Vlad Hondru, Radu~Tudor Ionescu, and Mubarak Shah.
\newblock Diffusion models in vision: A survey.
\newblock {\em IEEE Transactions on Pattern Analysis and Machine Intelligence}, 45(9):10850--10869, 2023.

\bibitem{ronneberger2015u}
Olaf Ronneberger, Philipp Fischer, and Thomas Brox.
\newblock U-net: Convolutional networks for biomedical image segmentation.
\newblock In {\em International Conference on Medical image computing and computer-assisted intervention}, pages 234--241. Springer, 2015.

\bibitem{Griffin1984}
Daniel~W. Griffin and Jae~S. Lim.
\newblock {Signal Estimation from Modified Short-Time Fourier Transform}.
\newblock {\em IEEE Trans. Acoust.}, ASSP-32(2):236--243, 1984.

\bibitem{khan2010geysers}
M~Ali Khan and J~Truschel.
\newblock The geysers geothermal field, an injection success story.
\newblock {\em GRC Trans}, 34:1239--1242, 2010.

\bibitem{zhu2019phasenet}
Weiqiang Zhu and Gregory~C Beroza.
\newblock Phasenet: a deep-neural-network-based seismic arrival-time picking method.
\newblock {\em Geophysical Journal International}, 216(1):261--273, 2019.

\bibitem{Nayak2018}
Avinash Nayak, Taka'aki Taira, Douglas~S. Dreger, and Roland Gritto.
\newblock {Empirical Green's tensor retrieved from ambient noise cross-correlations at The Geysers geothermal field, Northern California}.
\newblock {\em Geophys. J. Int.}, 213(1):340--369, 2018.

\bibitem{Lin2018}
Guoqing Lin and Bateer Wu.
\newblock {Seismic velocity structure and characteristics of induced seismicity at the Geysers Geothermal Field, eastern California}.
\newblock {\em Geothermics}, 71(September 2017):225--233, 2018.

\bibitem{Gritto2022}
Roland Gritto, Steve~P Jarpe, and David~L Alumbaugh.
\newblock {New Large-Scale Passive Seismic Monitoring at The Geysers Geothermal Reservoir, CA, USA}.
\newblock {\em Stanford Geotherm. Work.}, pages 1--11, 2022.

\end{thebibliography}

\end{document}